# Dynamics of proteins: Light scattering study of dilute and dense colloidal suspensions of eye lens homogenates

A. Giannopoulou,[1,2] A. J. Aletras,[3] N. Pharmakakis,[4] G. N. Papatheodorou,[1] and S. N. Yannopoulos[1,]*

[1] *Foundation for Research and Technology Hellas – Institute of Chemical Engineering and High Temperature Chemical Processes (FORTH / ICE-HT), P.O. Box 1414, GR-26504, Rio-Patras, Greece*
[2] *Department of Pharmacy, University of Patras, GR-26504, Rio-Patras, Greece*
[3] *Laboratory of Biochemistry, Department of Chemistry, University of Patras, GR-26504, Rio-Patras, Greece*
[4] *School of Medicine, Department of Ophthalmology, University of Patras, GR-26504, Rio-Patras, Greece*

**Abstract**
We report a dynamic light scattering study on protein suspensions of bovine lens homogenates at conditions (pH and ionic strength) similar to the physiological ones. Light scattering data were collected at two temperatures, 20 $^o$C and 37 $^o$C, over a wide range of concentrations from the very dilute limit up to the dense regime approaching to the physiological lens concentration. A comparison with experimental data from intact bovine lenses was advanced revealing differences between dispersions and lenses at similar concentrations. In the dilute regime two scattering entities were detected and identified with the long-time, self-diffusion modes of α-crystallins and their aggregates, which naturally exist in lens nucleus. Upon increasing protein concentration significant changes in time correlation function were observed starting at ~75 mg ml$^{-1}$ where a new mode originating from collective diffusive motions becomes visible. Self-diffusion coefficients are temperature insensitive, whereas the collective diffusion coefficient depends strongly on temperature revealing a reduction of the net repulsive interparticle forces with lowering temperature. While there are no rigorous theoretical approaches on particle diffusion properties for multi-component, non-ideal hard-sphere, polydispersed systems, as the suspensions studied here, a discussion of the volume fraction dependence of the long-time, self-diffusion coefficient in the context of existing theoretical approaches was undertaken. This study is purported to provide some insight into the complex light scattering pattern of intact lenses and the interactions between the constituent proteins that are responsible for lens transparency. This would lead to understand basic mechanisms of specific protein interactions that lead to lens opacification (cataract) under pathological conditions.

---

\* Corresponding author. E-mail: sny@iceht.forth.gr.

**I.    INTRODUCTION**

The dynamics of dispersions of colloidal particles has been the subject on intensive theoretical and experimental (mainly by means of light scattering) investigations over the last two decades [1]. Most of the studies have been focused on simple, hard sphere colloidal systems where both theoretical modeling and experimental implementation are possible. The accurate characterization of the interactions within such dispersions is useful since it can serve as the starting point to understand the behavior of more complex systems, which (i) depart from hard sphere behavior, (ii) exhibit appreciable polydispersity, and (iii) are of multi-component nature with highly asymmetric particle sizes. Biological systems, mainly protein dispersions, fit within the behavior of such complex dispersions and hence their study is useful to elucidate new aspects in the physics of colloids as well as to understand basic mechanisms of tissue functions and the routes to their degradation under pathological conditions.

Studies on structure and dynamics of protein dispersions are interesting subjects of research [2]. Proteins can self-assemble as a function of temperature, pH, ionic strength, concentration, purity, pressure, etc. Even small amounts of aggregates can significantly alter the molecular structure that can modify protein function. Monitoring and understanding these effects are fundamental for understanding the relation of the molecular mechanisms that lead to proteins self-assembly and the effect that this could have to cellular function for the onset or the progress of a related disease.

Perhaps, the most characteristic example of a mammalian tissue that exhibits complex colloidal behavior is the ocular lens. Lens proteins, called crystallins, are the major macromolecular content of the ocular lens. Apparently, their interactions play a decisive role in the underlying lens transparency. Lens opacification, usually referred to as cataract, reflects changes in interactions between crystallins and in particular are associated with changes in the short-range structural order, which has been considered as the most important factor for lens transparency [3-5]. Major attempts to account for lens transparency include, the long-range order or paracrystalline state theory of Trokel [3(a)], the short-range order theory [3(b,c), 4], the random fluctuation theory [5(a),(b)] and the gelation theory [5(c)]. Therefore, a thorough understanding of the molecular mechanisms of lens opacification, signifying the onset of cataract, inevitably calls for an appreciation of the interactions between protein "particles" and the aggregation processes that take place upon ageing and/or in the presence of other pathogenic factors.

In this paper, we present a systematic DLS study of whole lens homogenate dispersions over a wide range of concentrations. Lens homogenate dispersions have the advantage of being stable, avoiding crystallization, at very high concentrations. Our main aim is to gain some insight into the interparticle interactions by studying the diffusive motions or protein "particles", which will be a valuable basis for comprehending the complex relaxational pattern of the protein dynamics in the intact lens. This is the prerequisite for understanding the protein condensation processes that are responsible for lens opacification (such as cataract) and for developing a reliable methodology for *in vivo*, non-invasive diagnosis of ocular diseases.

The paper is organized as follows. Section II presents a brief summary on lens crystallins and light scattering from their suspensions. Section III contains the experimental informaiton including material preparation and DLS apparatus details. The data analysis procedures and the results obtained are presented in Sec. IV. Section IV is divided in three subsections. The first deals with data analysis details and the role of the various scattering entities in the scattering functions. In the second we provide a discussion on the identification of the relaxation modes and the role of size and optical polydispersity. The third subsection contains a detailed discussion about the volume fraction dependence of the collective diffusion coefficient and in particular about the long-time self-diffusion coefficient where a thorough comparison between our experimental data and exiting theoretical approaches is advanced. Finally, the most important conclusions drawn from the present study are summarized in Sec. V.

## II. BRIEF BACKGROUND ON LENS PROTEINS AND THEIR LIGHT SCATTERING STUDIES

Experimental techniques, such as dynamic light scattering (DLS), are ideally suitable for monitoring protein self-assembly processes and quantifying intermolecular interactions [6]. DLS has amply utilized in the past for *in vivo* [7] as well as *in vitro* [8] studies of intact mammalian lenses for this purpose. However, what has emerged from such studies is that the relaxation pattern of the ultra dense, gel-like eye lens cytoplasm is too complex in order to be reliably and uniquely rationalized in terms of the diffusive motions of the constituent protein "particles" or scattering elements [7, 8]. This obstacle can be overcome by a hierarchical approach where the lens proteins are isolated and studies of their dilute, semi-dilute and dense suspensionsare undertaken [9-11].

Eye lens is a protein/water dense suspension where crystallins' mass can reach up to 70% of the total tissue mass [12]. Three major types of crystallins are found in mammalian lenses, i.e. the α-, β-, and γ-crystallins. α-crystallins are large oligomers (40 to 60 particles) with molecular weight ($M_w$) in the range 800 – 1200 kDalton. β-crystallins are oligomers, either trimers ($β_L$) $M_w ≈ 60$ kDalton or octamers ($β_H$) with $M_w ≈ 160$ kDalton. γ-crystallins are monomers with $M_w ≈ 20$ kDalton. The relative concentrations, in wt%, of α-, $β_L$-, $β_H$- and γ-crystallins in the mammalian fiber cells are 45/20-25/10, respectively [12(b)], although this proportion varies with species as well as with location within a single lens. Given that the light scattering intensity in a colloidal dispersion of particles with different masses scales approximately with their $M_w$, it is obvious that α-crystallins features will dominate over the other two types of proteins in DLS.

DLS studies of crystallins' suspensions have been reported in the past by several authors [9-11]. Studies in suspensions can be distinguished in two main classes: those where a the total lens content (lens homogenate) is diluted and hence the analogy of the three types of crystallins in the suspension is preserved similar to that of the intact lens [9(a)], and those where the various species, usually the α- and γ-crystallins, are isolated and studied separately [9(b)-9(f), 10, 11]. Even in the sole work of lens homogenate studies [9(a)], lens proteins were extracted from the periphery of the lens (the cortex), where γ- crystallins are largely absent [12]. Although α-crystallins dominate in DLS, as mentioned above, and thus the dynamics of homogenate suspensions are dominated by the dynamics of this species, the study of homogenate suspensions is important in its own right due to the specific role of γ-crystallins that are considered to play a major role in cataract formation (or phase separation of the lens cytoplasm) due to their attractive interactions. Indeed, Takemoto *et al.* [13(a)] have recently demonstrated that during aging there is a decrease in the ability α-crystallins to bind to γ-crystallins. This alteration in protein interactions can perturb the formation of short-range order and hence lens transparency.

## III. EXPERIMENTAL
### A. Preparation of protein suspensions

Fresh young bovine eyes (3 months old) were obtained from local meat companies. Lenses were carefully released and after removing their capsules, they were mechanically homogenized (in a mortar) and extracted in a 190 mM phosphate buffer, pH: 7.3 (~10 ml / lens). Sodium azide 0.02 % w/w was added in the extract to prevent bacterial growth. Thorough mechanical homogenization took place in a mortar; this method and has been preferred over the handling of the lens in a homogenizer which, in general, develop high temperatures that can cause protein denaturation and insolubility. Dust, insoluble protein aggregates, membrane fragments of lens cells, as well as other insoluble materials were removed by centrifugation of the extract at 20,000 g for 1 h using a superspeed refrigerated centrifugation (Sorval RC-5). The procedure took place at 10 $^{o}$C, in order to avoid causing damage to the proteins because of the excess heat due to centrifugation. The clear supernatant was drawn with a sterile pipette and concentrated at first by ultrafiltration, using an Amicon cell and YM-10 membrane (cutoff ~10kD), and then under vacuum, using a dialysis tube with $M_w$ cutoff of ~10kD in an appropriate vacuum flask, to obtain a final concentration of protein 400-500 mg ml$^{-1}$. The ultrafiltration membrane and the dialysis tube allowed the suspension to be concentrated by eluding water as well as salt, in a way so as to maintain fixed the ionic strength of the suspension.

The obtained highly concentrated extract was then used as a stock for further dilutions with the same buffer. Suspensions of concentrations, 1, 10, 25, 50, 75, 100, 150, 200, 300 and 400 mg (protein) ml$^{-1}$ were prepared in cylindrical pyrex tubes of various diameters depending on the volume of the suspension. To avoid the presence of dust particles, which can affect the DLS data, the dilute suspensions were filtered through 0.2 μm filters while the denser ones were centrifuged for few minutes. The suspensions were equilibrated at 37$^{o}$C and 20$^{o}$C (accuracy: ±0.1$^{o}$C) with the aid of a temperature-controlled water bath.

Protein content of samples was determined by means of two methods. The Bradford protein assay method with bovine serum albumin (BSA) as a standard and the UV absorbance method. In Bradford method, an appropriate volume of each sample was mixed with Bradford reagent and after 2 min at room temperature the absorbance was measured at 595 nm, using a double beam UV/VIS spectrophotometer Cary 1E (Varian). The concentration of lens extract proteins in our samples was calculated using a standard curve constructed by suspensions of different concentrations of BSA. In the UV absorbance method, the protein sample was diluted in the buffer and placed in a rectangular cell with an optical path of 1 cm. The absorbance of the sample was then measured at 280 nm following the method reported in Ref. [9(a)]. Although there might be some uncertainty concerning the absolute values of the absorption experiments, the relative concentrations of the protein suspensions can be considered as accurate.

**B.   Dynamic Light Scattering apparatus and data analysis**

Normalized intensity time correlation function $g^{(2)}(q,t) = \langle I(q,t)\, I(q,0) \rangle / \langle I(q,0) \rangle^2$, were measured at right angle on a broad time scale (from 10$^{-8}$ s to 10$^4$ s) using a full multiple tau digital correlator (ALV−5000/FAST) with 280 channels. The scattering wavevector $q = 4\pi n \sin(\theta/2)/\lambda_0$ depends on the scattering angle $\theta$, the laser wavelength $\lambda_0$, and the refractive index of the medium *n*. The majority of measurements were conducted at right angle while selected data were also accumulated at various angles in order to check the diffusive character of the observed relaxation processes. The light source was an Ar$^+$ ion laser (Spectra Physics 2020) operating at 488 nm with a stabilized power of about 20 mW. The scattered light was collected by a single mode optical fiber mounted on a homemade goniometer for easy manipulation of the angular dependence.

Under the assumption of homodyne conditions the desired normalized electric-field time auto-correlation function $g^{(1)}(q, t)$ is related to the experimentally recorded function $g^{(2)}(q, t)$ through the Siegert relation [14]:

$$g^{(2)}(q,t) = B\,[1 + f^*\,|\,g^{(1)}(q,t)\,|^2\,] \qquad (1)$$

where $B$ describes the long delay time behavior of $g^{(2)}(q, t)$ and $f^*$ represents an instrumental factor obtained experimentally from measurements of a dilute polystyrene/toluene suspension. In our case, the optical fiber collection results in $f^* \approx 0.95$.

The electric-filed time correlation function $g^{(1)}(t)$ – for simplicity we drop the $q$-dependence in the following – was analyzed as a weighted sum of independent exponential contributions, i.e.:

$$g^{(1)}(t) = \int L(\tau) \exp(-t/\tau) \, d\tau = \int L(\ln \tau) \exp(-t/\tau) \, d\ln \tau \qquad (2)$$

where the second equality is the logarithmic representation of the relaxation times. The distribution of relaxation times $L(\ln \tau)$ was obtained by the inverse Laplace transformation (ILT) of $g^{(1)}(q, t)$ using the CONTIN algorithm [15].

Alternatively, $g^{(1)}(q, t)$ was fitted with a sum of stretched exponential functions; the summation index $i$ depends upon the protein concentration $c$,

$$g^{(1)}(t) = \sum_i A_i \exp[-(t/\tau_i)^{\beta_i}], \qquad i: 1, 2, \ldots \qquad (3)$$

where $A_i$ is the amplitude (zero-time intercept) of the $i^{th}$ decay step, and $\beta_i$ is the corresponding stretching exponent which is characteristic of the breadth of the distribution of the relaxation times and assumes values in the interval [0, 1]. The indices $i$ denote the number of the decay modes involved.

## IV. RESULTS AND DISCUSSION
### A. Data analysis details and the role of the various scattering entities

Representative intensity time correlation functions for various concentrations of the lens cytoplasmic homogenates at 20 $^{o}$C are shown in Fig. 1. Open symbols correspond to the experimental data, while solid lines passing through the symbols represent the best-fit results obtained with the use of the CONTIN algorithm (Eq. 2). It is obvious that when the concentration of the crystallins increases the decay of the correlation function becomes progressively broader implying that new relaxation modes appear. At the highest concentration, which is near the physiological one of the bovine lens nucleus, the intensity correlation function (solid circles) displays a decay behavior that is faster than exponential, at short times, while ripples or oscillations also appear. This is a common behavior in DLS studies of gels although ripples are rarely reported because they are considered as experimental artifacts of the scattered light. On the other hand, it has been shown [16] that a wealth of information can be obtained analyzing such oscillations. For comparison we have added in Fig. 1 the experimental correlation functions of the intact bovine lens recorded from the lens nucleus at 37 $^{o}$C (dashed-dotted line) and at 20 $^{o}$C (solid line) as well as from the lens cortex (dashed line) regions.

The data shown in Fig. 1 reveal a strong concentration dependence of the correlation functions. Indeed, the increase of concentration causes a gradual slowing-down of the characteristic times as well as the appearance of slower modes whose origin will be discussed below. The data of the intact bovine lens reveal a very strong temperature dependence of the correlation function of the intact lens between 20 $^{o}$C and 37 $^{o}$C. Changes pertain both to relaxation times and amplitudes of the fast (~$10^{-1}$ ms) and slow (~$10^{1}$ ms) processes of $g^{(2)}(t)$. Such drastic changes are anticipated in view of the onset of the lens cytoplasm phase separation or the so-called "cold cataract" effect and the concomitant opacification that occurs in the intact bovine lens when lowering the temperature below 20 $^{o}$C [7(b), 8(e)]. However, the corresponding temperature changes [17] of the lens homogenate are much milder pointing to the importance of the short range order of lens proteins in the fiber cells of the lens. Another possible source of the increased scattering in the lens is related to fluctuations in the orientation of optically anisotropic protein "particles" in the fiber cells. This can cause intrinsic and form birefringence. The former originates from alignment of large anisotropic structures and the latter arises from the fact that geometrically anisotropic structures are embedded in a medium with different refractive index [18].

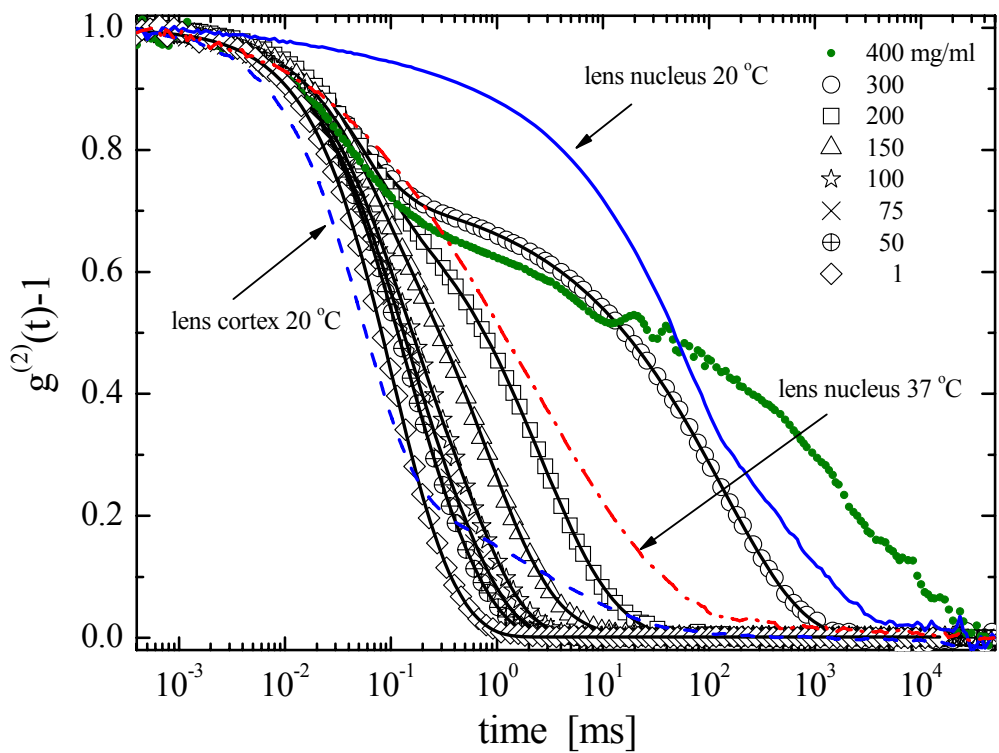

**Fig. 1:** *Normalized intensity auto-correlation functions of bovine lens protein dispersions for various concentrations as shown in the legend. For comparison, the experimental correlation functions of the intact bovine lens recorded from the nuclear at 37 $^oC$ (dashed-dotted line) and at 20 $^oC$ (solid line) as well as from the cortex (dashed line) regions, are shown. Solid lines passing through the data points (open symbols) represent the best fit results obtained with the aid of Eq. (2) and (3). The fit results of both equations are indistinguishable.*

Before proceeding to the analysis and the interpretation of the experimental data it would be instructive to recapitulate the relative role of the various types of crystallin proteins in the intensity correlation functions. Reckoning on the fact that the scattering strength of each species is proportional to the product $c \times M_w$ and taking into account the relative fraction of each protein ($\alpha$:$\beta$:$\gamma$ = 45:42:13) it can be shown that the total scattered intensity emerging from $\beta$- and $\gamma$- species amount to only 14% of the scattered intensity of the $\alpha$-crystallins. In practice, the high molecular weight $\beta_H$-crystallins contribute the largest part of this 14%, while the low molecular weight $\beta_L$- and $\gamma$-crystallins play a role in the total volume fraction as well as in the interparticle interactions. It is interesting to notice that the total scattered intensity of crystallins' suspension is an increasing function of concentration in dilute suspensions. This suggests that at high concentrations, i.e. comparable to those in the lens nucleus (300-400 mg ml$^{-1}$) the increasing scattering of light would lead to opacification. However, this is true neither for the lens nucleus nor for dense homogenate suspensions [4(a)] where it has been found that the scattered light intensity increases up to $c \approx$ 100-120 mg ml$^{-1}$ followed by a drastic decrease at higher concentrations.

This effect was also observed in this work where the transmitted light intensity has been recorded and is shown in Fig. 2. As this figure reveals, the normalized transmitted light ($I_t / I_0$, i.e. transmitted over incident intensity) through the crystallins' suspensions decreases reaching a minimum around $c \approx 100$ mg ml$^{-1}$ while then increases to higher levels indicating the enhanced transparency of the suspensions at concentrations near the physiological one. The non-monotonic behavior of transparency as a function of concentration has been attributed to the fact that the suspension becomes progressively more homogeneous due to the special close packing of the lens proteins that fill-up the space avoiding crystalline-like order. As a result the refractive index fluctuations are minimized and ultimately opacification is related to the local changes in the refractive index within the lens [3, 5].

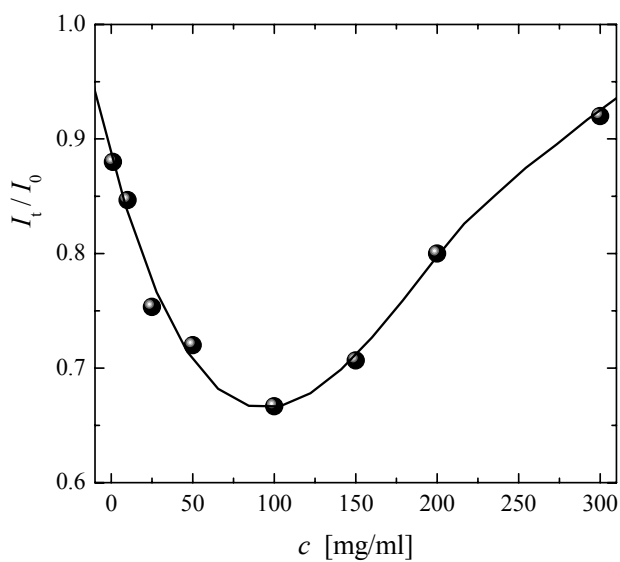

**Fig. 2:** *Concentration dependence of the normalized transmitted light intensity ($I_t$) through the suspensions of crystallins homogenates at 20 ºC. $I_0$ denotes the incident laser intensity. The line is drawn as a guide to the eye.*

Closing this subsection we would like to notice two points. First, irrespectively of the concentration, no special *q*-dependence of the total scattered intensity was observed. This is expected in view of the form factor of the crystallins and the fact that we are at the $qR_g \ll 1$ limit (Guinier regime), where $R_g$ is the radius of gyration of the protein "particle". The absence of appreciable *q*-dependence of the scattered intensity implies also the lack of strong correlations in spatial ordering of the protein suspensions. Second, the transmitted intensity reduction at moderate concentrations is not so significant so as to produce appreciable multiple scattering effects that could severely affect the analysis of the correlation functions.

**B.    Identification of the relaxation modes: the role of polydispersity**

The analysis of the experimental data showed that even for the most dilute suspension ($c$ = 1 mg ml$^{-1}$) the correlation function could not be satisfactorily fitted with a simple exponential decay. Even the use of a cumulants analysis retaining up to the second moment [see inset in Fig. 3(a)], i.e. taking into consideration polydispersity, does not lead to a satisfactory fit to the experimental data of the dilute suspension. Similarly poor results were found in an attempt to fit these same data with a stretched exponential function [Eq. (3)] that accounts for a broader size distribution for values of the stretching exponent $\beta < 1$. The application of the ILT method (using CONTIN) showed, see Fig. 3(a), that even at this very low concentration there is a bimodal particle size distribution with relaxation times that differ by almost half a decade.

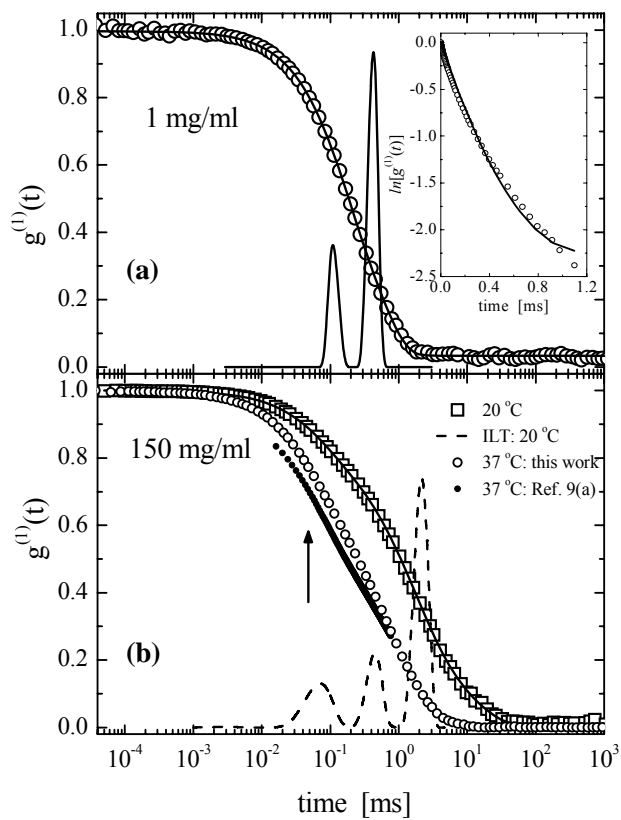

**Fig. 3:** *Representative analyses of the time correlation functions for $c = 1$ mg ml$^{-1}$ (a) and $c = 150$ mg ml$^{-1}$ at 20 °C. The inset in (a) illustrates an attempt to fit the very dilute dispersion with the aid of cumulants analysis taking into account polydispersity. For details see text. Part (b) contains for comparison our data at 37 °C (open circles) and the corresponding time-correlation function from cortex lens homogenates from Ref [9(a)] (closed circles). The arrow denotes the relaxation time estimated in the analysis of Ref. [9(a)].*

In the only existing study of lens homogenate protein suspensions [9(a)] the corresponding time correlation function of the dilute suspension was considered as a single exponential one obviously due to the inadequate experimental resolution and the limited time domain accessible of the correlator used in that work. This is more evident when considering Fig. 3(b) where the experimental data and the ILT analysis of a moderately dense protein suspension of concentration $c = 150$ mg ml$^{-1}$ are illustrated. The solid circles denote the experimental time correlation function taken from Ref. [9(a)] for the same concentration (150 mg ml$^{-1}$) and at the physiological temperature (37 °C); our data at the same temperature are shown for comparison. Obviously, the lack of both the short-time plateau as well as the long-time baseline from the data of [9(a)] (filled circles) imply that the features of protein dynamics in dilute and dense suspensions reported therein were not accurately established. As a consequence of the limited range of the intensity correlation functions in [9(a)], only one fast decay time was estimated and a slow one was hypothesized based on the fact that the long-time baseline did not decay to zero. In contrast, the broad time scale being recorded in the present work makes it possible to correctly determine the number of relaxation modes as shown in Fig. 3(b). The vertical arrow denotes the relaxation time estimated in [9(a)] for the depicted time correlation function.

The existence of two decays steps in the time correlation function of the dilute homogenate suspensions is also supported by the result of Fig. 4(a) where the data are shown in a modified plot of $\log[-\log(g^{(1)}(t))]$ vs. $\log t$. The slope of the data in this modified plot represents the stretching exponent $\beta$ (see Eq. 3); which in the case of single exponential decays should be unity. The slope=1 is shown in Fig. 4(a) alongside with the data of the concentration $c = 1$ mg ml$^{-1}$. Obviously, a single exponential decay is insufficient to fit the data even at this very dilute regime. Using this way of plotting relaxational data can also help justifying the presence of the increasing number of distribution peaks of $L(\ln \tau)$ at denser suspensions. For instance, Fig. 4(b) illustrates another example of the data that correspond to the concentration $c = 150$ mg ml$^{-1}$ and the corresponding ILT distribution. There is an noticeable slope change at a point located just below $10^{-1}$ ms where the fast relaxation time regime reveals a quasi-single exponential decay (slope: 0.9) while the long relaxation part ($t > 10^{-1}$ ms) is characterized by a considerably small value of the stretching exponent, i.e. $\beta = 0.54$. It is therefore clear why the ILT distribution [$L(\ln \tau)$ curve] conceals more than one peak in the long time decay step.

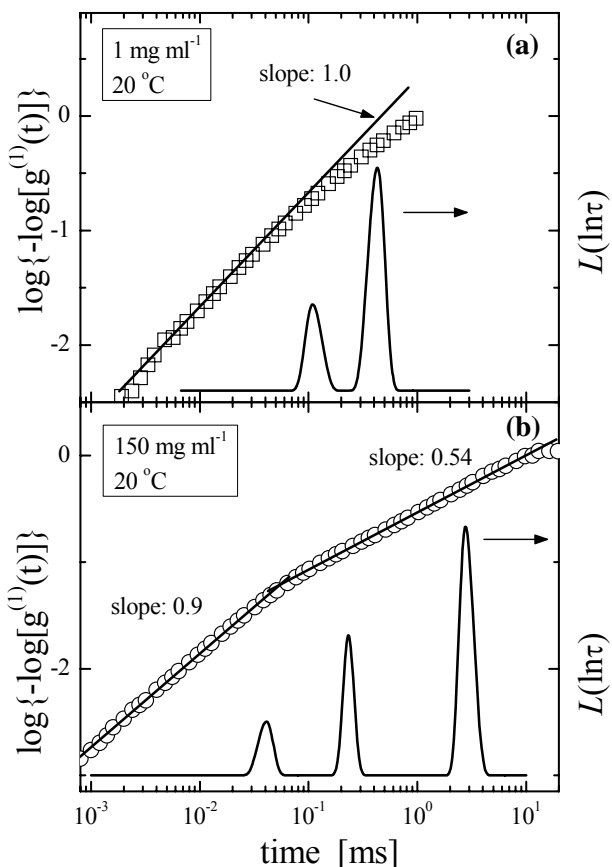

**Fig. 4:** *Modified logarithmic representation of the time correlation functions for c = 1 mg ml$^{-1}$ (a) and c = 150 mg ml$^{-1}$ at 20 °C demonstrating the deviation of the experimental functions from the simple exponential form in the dilute dispersion (a), and the justification of the existence of three relaxation modes in the semi-dilute dispersion (b).*

The concentration dependence of the distributions of the relaxation times obtained with the aid of CONTIN is shown in Fig. 5. This figure reveals that the dynamics behave smoothly up to $c$ = 75 mg ml$^{-1}$; the two ILT peaks shift to longer times as $c$ increases. At this concentration an important change takes place with the onset of a new – faster than the existing ones – relaxation mode. The observation that the relaxation pattern becomes more complicated has also been reported in studies of pure α-crystallin suspensions [10(c)]. The relaxation time of the new (fast) mode exhibits opposing trend as a function of $c$, namely, it becomes faster with increasing concentration. This behavior suggests that the two modes, which are already observed at very low protein concentrations, can be identified with the self-diffusive motions from where the self-diffusion coefficients $D_{S-1}$ and $D_{S-2}$ of α-crystallins and high molecular weight aggregates of α-crystallins (HMα), respectively can be estimated. The fact that these slow modes are visible in DLS at very low concentrations can be attributed to the polydispersed nature of these scattering elements. Indeed, as mentioned above, molecular weights in the range 800 – 1200 kDalton are frequently considered for α-crystallins; however, values ranging from 280 kDa to above 40,000 kDa have also been reported [19(a), (b)]. Moreover, the ionic strength of the suspensions seems to strongly affect the $M_w$ and the polydispersity; at high ionic strength, as is the present case, proteins with higher $M_w$ have been found [19(c)].

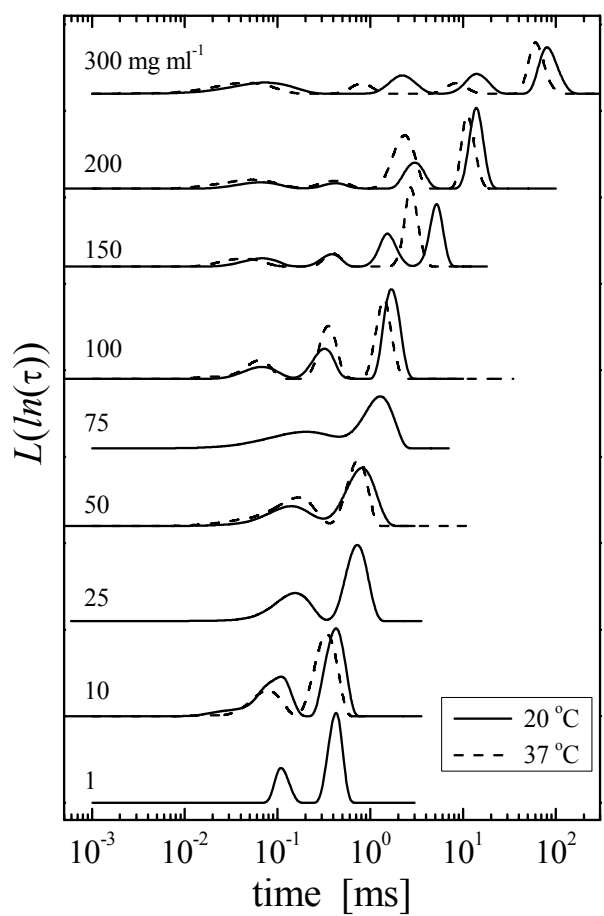

**Fig. 5:** *ILT distributions as a function of protein concentration of lens homogenate suspensions at 20 $^o$C (solid lines) and 37 $^o$C (dashed lines).*

In a detailed investigation of hydrodynamic and conformational properties of lens proteins [19(d)], it was reported that α-crystallins (in their native form) and HMα possess considerable size and charge heterogeneities in their native structure and subunit polypeptides, respectively. This finding accounts for the observability of the two modes in the ILT distributions for low concentrations, as shown in Fig. 5, which can be associated with the diffusion modes of α and HMα "particles". An estimation of the $M_w$ of the HMα scattering elements can proceed considering that for such three-dimensional, spherical-like objects [4(b)] the $M_w$ is proportional to the $R^3$, where $R$ denotes the particle radius. Because the ratio of the diffusion coefficients $D_{S-1}/D_{S-2}$ is of about 3.67 we expect that $M_w^{HM\alpha}/M_w^{\alpha} \approx 50$, which results in $M_w^{HM\alpha} \approx 40,000\ kDa$ taking into account that $M_w^{\alpha} \approx 800\ kDa$. However, knowing that the dry volume of the HMα particles is of about 30% of the hydrated volume [4(c)], we expect that $M_w^{HM\alpha}$ might not exceed ~12,000 *kDalton*. This is a reasonable value that falls within the $M_w$ range of HMα "particles" [19 (a), (b)].

At concentrations higher than 150 mg ml$^{-1}$ we observe the appearance of a fourth slow mode, (Fig. 5). Four modes have been identified in studies of intact lenses (see for example Refs. [8(b), (e)]). Their identification became possibly lately with the advent of logarithmic correlators that are able to collect data over a very broad time scale spanning more than eight decades in time. The slow mode is reminiscent to the very slow structural relaxation modes observed in structural glasses termed as long-range density fluctuations. It is thus reasonable to assume that in highly dense colloidal suspensions, approaching their glass transition, analogous modes may also exist.

The fast mode that appears at fast times near the concentration $c \approx 75$ mg ml$^{-1}$ can possibly be associated with the collective diffusion motions (collective or mutual diffusion coefficient $D_C$) resulting from collective concentration fluctuations which are controlled by the osmotic pressure of the suspension [1(a), (c), (d)]. It is worth noting here that in a previous DLS study of lens homogenates [9(a)] only one relaxation mode was identified which was attributed to the mutual or collective diffusion. The limited range of the intensity correlation functions in [9(a)] led the authors to hypothesize the existence of a slowly diffusing mode, associated with the α "particles", based on the fact that the long-time baseline did not decay to zero. On the other hand, HMα "particles" have only been observed in lens nucleus, but not in α-crystallins' suspensions extracted from lens cortex [19(e)]. The appearance of the collective mode in our data above some certain protein concentration seems to contradict the observation of [9(a)] where this mode is "visible" in DLS even at the very dilute suspensions. This apparent contradiction can be elucidated invoking the consequences of polydispersity of the scattering elements, present in whole lens homogenates, on the relaxation amplitudes of the relaxation modes.

The effect of polydispersity in DLS is rather complicated; one has to distinguish between *size*, and *optical* (or scattering-power) polydispersity [1(c), 20]. Although the impact of the two limiting cases of these two types of polydispersity, i.e. only size or only optical polydispersity, on the time correlation function is rather well understood, the presence of both types is highly complicated [1(c), 20]. Size polydispersity is usually invoked in dilute suspensions and the initial slop of $ln[g^{(1)}(t)]$ is used to define the polydispersed diffusion coefficient whose oscillatory behavior as a function of the wavevector is adequately understood. On the other hand, optical polydispersity can be exploited to study experimentally different kinds of diffusion processes and hence its effects are discussed for concentrated suspensions where the interactions between Brownian particles are more important [1(c)]. In this case, both the collective dynamic structure factor $S_C(q,t)$, as well as the self-dynamic structure factor $S_S(q,t)$ are important. The time dependence of $S_C(q,t)$ describes the dynamics of the sinusoidal density fluctuations with wavelength $\Lambda = 2\pi/q$. Because such fluctuations involve the simultaneous movement of many particles $S_C(q,t)$ is related to collective phenomena. On the other hand, $S_S(q,t)$ describes the dynamics of a single particle, which is inevitably affected by the interactions with other particles. In this case, the polydispersed electric-field time correlation function is the sum of a self and a collective term with mode amplitudes (weighting factors) that depend upon the scattering amplitudes of the particles. In cases where the time scales for self and collective diffusion are sufficiently different they can both be determined through the same time correlation function. This is the case in our study at $c > 75$ mg ml$^{-1}$.

Theoretical works have focused on the relative contributions of the collective and self-diffusion processes. Optical polydispersity is simpler to cope with theoretically than size polydispersity although systems for which the theory applies exactly, i.e. particles of the same size that differ in refractive index only, are vary rare. The main effect of optical polydispersity, as mentioned above, is to introduce into the measured dynamic structure factor another slower term that describes single-particle motions. Optical polydispersity alone has no effect in a noninteracting system, i.e. in a very dilute suspension [21(c)], therefore this type of polydispersity cannot be considered as the main cause of the appearance of the slow modes of the lens homogenate suspensions at low concentrations presented in Fig. 5. Further, the effects of optical polydispersity become more evident in the hydrodynamic regime, $q << q_m$, where $2\pi/q_m$ is roughly the most probable interparticle spacing, and the effects of even a small extent of optical polydispersity can be rather easily detected. This supports the contribution of optical polydispersity in our high concentration suspension where the hydrodynamic interactions between the particles are more important. The relative amplitude of the collective and self-diffusion processes is a complicated function of the effects induced by the type of polydispersity present in the system [1(c), 20]. In the case of lens homogenates studied in this work there are presumably both size and optical polydispersity, as well as charge polydispersity due to the different charges carried by α-, β-, and γ-crystallins [12]. It is therefore reasonable to assume that the amplitude of the collective mode is severely suppressed at low protein concentrations due to the presence of the complicated polydispersity discussed above. Further, the proximity in the times scale of the collective and self-diffusion processes at low $c$ is also another factor that may account for the "invisibility" of the former.

Support to the aforementioned also comes from the fact that proteins used in the DLS study in [9(a)] were extracted from the cortex of the lens where one species of crystallins (γ-crystallins) is largely absent and high molecular weight aggregates of the α-crystallins (responsible for the slowest diffusive motion shown in Fig. 5, $D_{S-2}$) are also lacking. The lower degree of polydispersity present in lens homogenates produced by lens cortex proteins is also evident from the dashed curve shown in Fig. 1 that represents the time correlation function of the lens cortex at 20 °C. In this case, the amplitude of the fast decaying component (collective diffusion) is much larger than the corresponding amplitude of the homogenate suspensions and is almost equal (in the $g^{(1)}(t)$ representation) to the intensity the amplitude of the slow component related to self-diffusion. Finally, theoretical calculations have shown [1(c)] that the self-diffusion (slow) mode amplitude grows parallel to the increase of the degree of polydispersity and increases with increasing concentration as it also observed in our experimental study.

## C.  Volume fraction dependence of diffusion coefficients
### *1.  Remarks on interactions*

Figure 6 contains the volume fraction dependence of the various diffusion coefficients obtained from the analyses of the homogenate suspensions studied at 37 °C and 20 °C. Results from previous investigations on lens cortex homogenates [9(a)] and on α-crystallins [9(d)] suspensions are also shown for comparison. The exact transformation from $c$ to $\phi$ is a particularly important step in the analysis of the volume fraction dependence of the diffusion coefficient and especially in the case where a comparison of theoretical predictions and experimental data is attempted. While most experimental data of the self-diffusion coefficient of colloidal suspensions fit within theoretical predictions at high $\phi$, systematically lower values for $D_S$ are observed for proteins. In the absence of some (less probable) effects such as the enhancement of interactions due to the formation of transient oligomeric aggregates, the above discrepancy between experiment and theory may arise from the use of the dry specific volume $\upsilon_{sp}^{dry}$ in the calculation of $\phi$, i.e. $\phi = c \times \upsilon_{sp}^{dry}$. Neglecting the hydration contribution to the hydrodynamic volume is a severe approximation since this can be a considerable fraction of the excluded volume. In addition, the hydrodynamic volume is concentration dependent and this must also be taken into account. For an accurate transformation from $c$ to $\phi$ we used the experimental values of Ref. [4(c)] where the volume fraction dependence on protein concentration is given by $\phi/c \approx 1.953 - 1.72 \times c$. The last relation has been derived for α-crystallin dispersions and therefore its use for lens homogenates (mixtures of all types of crystallins) might entail some error. However, it is far more realistic than using the dry specific volume $\upsilon_{sp}^{dry}$ in the transformation $c \rightarrow \phi$.

Before discussing in detail the concentration dependence of the diffusion coefficients it would be instructive to briefly present some general issues concerning the interactions in colloidal suspensions [1]. It is known that in the absence of external forces the dynamics of colloidal particles suspended in a molecular solvent are governed by three types of interparticle interactions. These interactions involve solute-solvent, solute-solute, and solvent-solvent interactions. Solute-solvent interactions, known as Brownian forces, arise from random collisions between solute and solvent particles and determine primarily the diffusion mechanism. Solute-solute interactions are divided into direct ones – which can be repulsive (excluded volume and Coulombic interactions) or attractive (van der Waals) and indirect interactions. The latter arise because the motion of a solute molecule induces a fluid flow in the solvent that affects neighboring solute molecules. The effects on the diffusion coefficient caused by both direct (potential) and indirect (hydrodynamic) interactions depend on solute concentration.

An important parameter of dynamic light scattering experiments is related to the time scale at which the experiment is carried out in relation to the characteristic time scales of particle interactions. In the dilute case, non-interacting particles diffuse with a mean square displacement linear in time for $t \gg \tau_B$ where $\tau_B$ is the Brownian time, which sets the time scale for the average velocity relaxation of a colloidal sphere. Indirect (hydrodynamic) interactions propagate on a similar time scale, i.e. $\tau_H \approx \tau_B$. On the other hand, direct (potential) forces become effective on the interaction time scale $\tau_I$; this is the time scale of interactions between solute molecules. In an approximate estimation, $\tau_I$ can be considered as the time that a solute particle needs to diffuse a distance comparable to its own diameter and is approximately given by $\tau_I \approx R^2/D_0$, where $D_0$ is the $c \rightarrow 0$ magnitude of the diffusion coefficient, and hence lies within the sub-millisecond time regime. In our case, taking $R \approx 12$ nm [12(b)] for the radius of α-crystallins and using $D_0$ of the present work (see below) we obtain $\tau_I \approx 8\ \mu s$. From this result it is obvious that particle dynamics studied in our experiment pertain to the time regime where $\tau_B \ll \tau_I \ll t$. This is the long-time Brownian regime and the relevant diffusion coefficient is the *long-time self-diffusion coefficient $D_S^L$*. At shorter times, i.e. $\tau_B \ll t \ll \tau_I$, the motion is also diffusive, although faster due to the fact that particles do not "feel" the slowing down by the direct interactions of neighboring particles, being thus trapped in the transient "cage" formed by its neighbors. This is the short-time Brownian regime and the relevant diffusion coefficient is the *short-time self-diffusion coefficient $D_S^S$*. According to the aforementioned $D_S^L < D_S^S < D_0$. All these diffusion coefficients become equal in the absence of interactions or at $c \to 0$.

In view of the aforementioned discussion, Fig. 6 reveals the following. The collective diffusion motions becomes faster with increasing volume fraction which implies the dominance of repulsive interactions in the suspension. The slope of $D(c)$ vs. $c$ curve changes drastically when cooling the protein dispersions from 37 °C to 20 °C. The quantity measured here can be identified with the long-time collective diffusion coefficient. This mode is considered to describe the decay of concentration fluctuations associated to the average extension of a nearest neighbor cage [1(d)]. According to this definition, theoretical predictions have shown that the long-time collective mode cannot exist at low concentrations (i.e. for $\phi < 0.2$ for hard spheres) due to the weakening of the caging effect [1(d)] as our experimental data support. We cannot however exclude the possibility of having the contribution from the short-time collective diffusion coefficient in view of the very small difference (~6%) between $D_C^L$ and $D_C^S$ as recently predicted by calculations in dense hard-sphere systems [21].

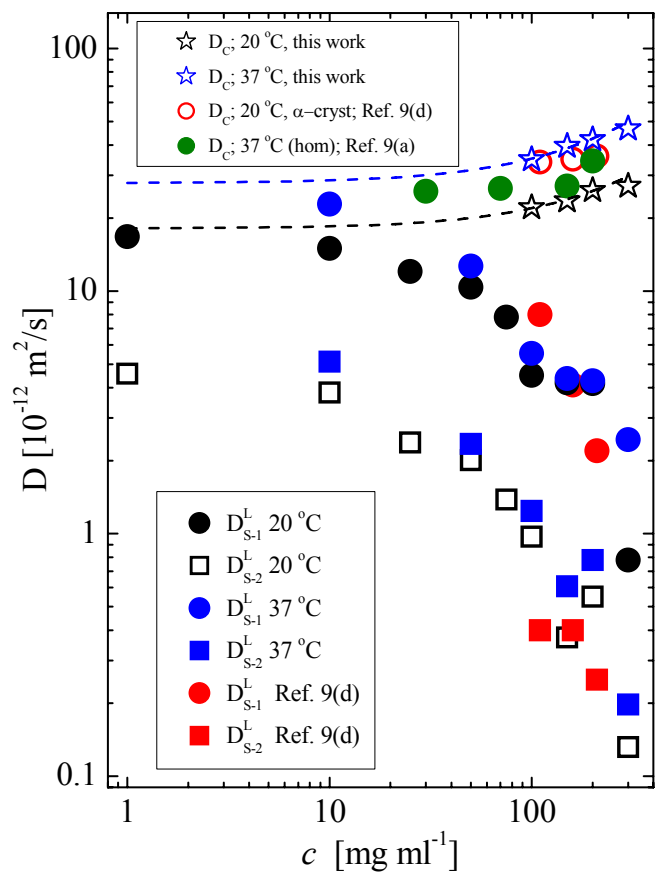

**Fig. 6:** *Double logarithmic representations of the concentration dependence of the long-time self-diffusion and collective diffusion coefficients of lens homogenate suspensions at 20 °C and 37 °C. The dashed line represent linear (in volume fraction) fits of the collective diffusion data extrapolated at the $c\to 0$ limit. Corresponding data from previous studies on lens cortex homogenates and α-crystallin dispersions are also included for comparison; see text for details.*

As regards the self-diffusion coefficient, $D_{S-1}^L$ and $D_{S-2}^L$ represent the long-time self-diffusion coefficients for α-crystallins and HMα particles, respectively. As is evident from Fig. 6, the concentration dependence of both $D_S^L$ is much stronger than that of the collective diffusion coefficient and exhibits a decrease with increasing *c*. On the contrary, the temperature dependence of $D_S^L$ is much weaker than that of $D_C$. This effect is analogous to the ionic strength dependence of the collective and self-diffusion coefficients [9(d)], where again $D_C$ exhibited appreciable ionic strength dependence while the $D_S^L$ vs. *c* curves were indistinguishable between two very different ionic strengths. Data for collective diffusion of lens homogenates at 37 °C and α-crystallins suspensions (Refs. [9(a), (d)]), as well as data for the self diffusion coefficient of α-crystallins suspensions (Ref. [9(d)]) fit well with the trends followed by our data.

## 2. Long-time self-diffusion coefficient

The above discussion shows that the diffusion coefficient of a solute particle depends on the concentration or volume fraction $\phi$ of the suspension because the strength of interactions also depends on concentration. Theories that deal with the volume fraction dependence of the diffusion coefficient, which where considered to be operative at the long-time regime, have appeared long time ago [22(a)-(g)]. However, this was later challenged [22(h)] as it was shown that the theories apply in the short-time regime, i.e. $\tau_B \ll t \ll \tau_I$. The volume fraction dependence of the long time self-diffusion coefficient in colloidal suspensions has over the last years been the focus of many theoretical and experimental works, as has been reviewed in [1(c), (d)], some of which will be briefly mentioned and their predictions will be compared with our data. More recently, studies have extended to include colloidal mixtures [23].

A simple, linear in volume fraction, relation has been suggested [22(g)] where for low volume fractions reads as:

$$D_S^L(\phi)/D_0 \approx 1 - 2.0972\,\phi\,. \tag{4}$$

A similar relation with slope 2.06 has been calculated in a subsequent paper [22(i)]. Experimental data at low volume fractions usually exhibit a higher slope than that predicted by the theories [24].

An important work on the correct description of $D_S^L(\phi)$ in neutral colloidal dispersions for high volume fractions was advanced by Medina-Noyola [25(a)], which was further developed by Brady [25(b)]. The model includes hydrodynamic interactions and seems to work for volume fractions up to $\phi \approx 0.5$. In this approach, it was suggested as a first approximation to decouple, at high volume fractions, hydrodynamic and direct interactions, i.e.

$$D_S^L \approx D_S^S\, D_S^H / D_0 \tag{5}$$

where $D_S^S$ is the short-time self-diffusion coefficient which incorporates hydrodynamic interactions, and $D_S^H$ is the long-time self-diffusion coefficient which neglects hydrodynamic interactions. It is generally considered that the assumptions made in [25(a)] lead to erroneous results at low $\phi$. In particular, using the approximation of Eq. (5) it can be estimated that the linear expansion coefficient [cf. Eq. (4)] assumes the value –3.831 [24(b)]. The application of the relation proposed in [25] requires the knowledge of the short-time self-diffusion coefficient. The analytical expression for $D_S^L(\phi)$ in terms of this theory and using analytical expression for $D_S^S$ and $D_S^H$ (for details see [24(b)]), acquires the form:

$$D_S^L(\phi)/D_0 = (1-\phi)^3 \, [1+(3/2)\phi + 2\phi^2 + 3\phi^3]^{-1}. \tag{6}$$

The advantage of this relation is that it does not involve any adjustable parameter. This relation is plotted in Fig. 7 a solid line and seems to describe rather well the data for $D_{S-1}^L(\phi)$. The dashed straight line represents the linear fit of $D_{S-1}^L(\phi)$ at low $\phi$ values ($\phi$<0.1) resulting in a slope of about -4. Interestingly, this value in near the slope (–3.831) predicted by this theory as described above.

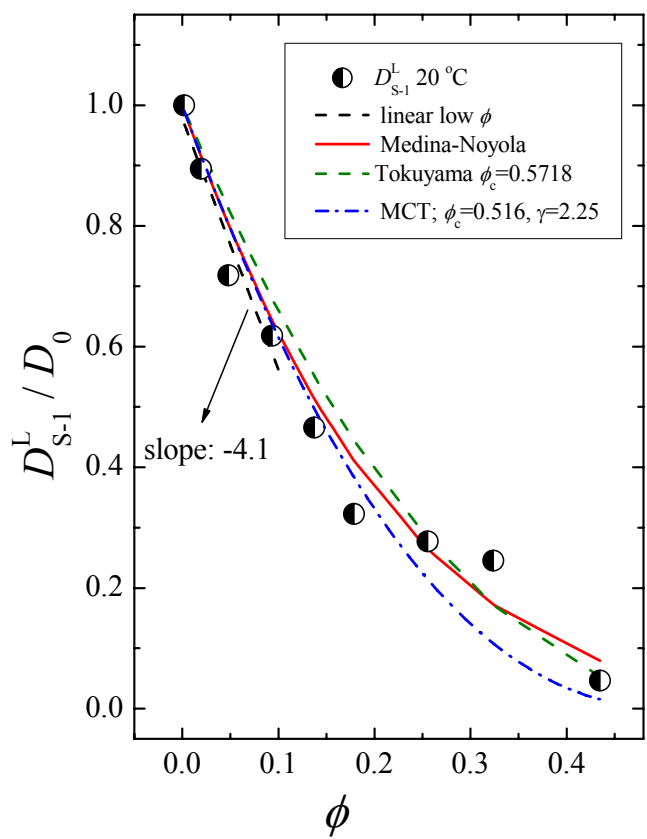

**Fig. 7:** *Volume fraction dependence of the long-time self-diffusion coefficient at 20 $^o$C. The various lines represent fits of the experimental data with the aid of theoretical approaches as described in the text.*

From another viewpoint, slow dynamics of hard sphere suspensions have been studied by Tokuyama and co-workers taking into account both hydrodynamic and direct interactions [26]. It was shown that both short- and long-range hydrodynamic interactions between particles play an important role in the dynamics of concentrated hard-sphere suspensions. The volume fraction dependence of $D_S^L(\phi)$ was described in the following equation:

$$\frac{D_S^L}{D_0} = \frac{\frac{D_S^S}{D_0}(1-9\phi/32)}{1+\varepsilon\frac{D_S^S}{D_0}\left(\frac{\phi}{\phi_c}\right)\left(1-\frac{\phi}{\phi_c}\right)^{-2}} \qquad (7)$$

where the short-time self-diffusion coefficient $D_S^S$ is given by Eq. 11 in [26(a)]. $\phi_c \approx 0.5718$ denotes the volume fraction at the colloidal glass transition for hard monodispersed spheres, and $\varepsilon$ is parameter introduced at the latter stages of the formulation of this theory, which can be determined by fitting Eq. (7) to the experimental data. For neutral hard spheres $\varepsilon = 1$, while this parameter increases appreciably higher than unity in the case that some soft behavior is exhibited by the suspended particles [26(f)]. The above equation provides a parameterless description of the experimental data of hard sphere colloidal suspension if $\phi_c \approx 0.5718$ and $\varepsilon = 1$. The dashed line in Fig. 7 represents the result of Eq. (7) for these conditions.

Knowing that suspensions of crystallins studied in this work contain particles of different sizes and exhibit appreciable polydispersity for each size we have attempted to fit the data for $D_S^L(\phi)$ with Tokuyama's equation using $\phi_c$ and $\varepsilon$ as free fitting parameters. To increase the reliability of the fit we have included also related data from a polydispersed soft sphere system obtained by Brownian dynamics simulations near the glass transition [27(a)]. The fit result is shown in Fig. (8) by the solid line for $\phi_c \approx 0.594 \pm 0.001$ and $\varepsilon = 2.39$, which implies some degree of softness in the protein "particles" as was expected. More recently, based on the fact that experimental and theoretical data show no exact singular behavior at $\phi_c$, the mean field theory of Tokuyama has been advanced to account for the avoidance of singularity of $D_S^L(\phi)$ [26(f)].

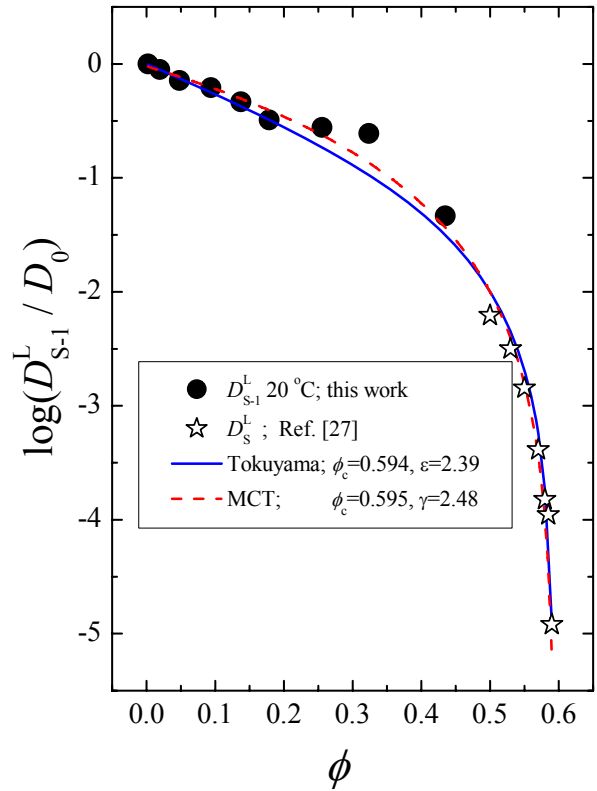

**Fig. 8:** *Logarithmic representation of the volume fraction dependence of the long-time self-diffusion coefficient at 20 ºC. Available data from recent simulations for hard sphere systems near the glass transition point of this system have also been included to facilitate the fit with the two theoretical models as described in the text.*

Predictions on slow dynamics of colloidal suspensions are also provided by the mode coupling theory (MCT) [28]. The prediction of MCT on the long-time self-diffusion coefficient close to the glass transition is a power law asymptote given by the following relation:

$$D_S^L(\phi) \propto (1 - \phi/\phi_c)^\gamma \qquad (8)$$

with the critical exponent $\gamma = 2.6$. Fitting our data, in conjunction to the data of Ref. [27(a)], with the equation predicted by MCT we obtain $\phi_c \approx 0.595 \pm 0.001$ and $\gamma = 2.49 \pm 0.12$; dashed line in Fig. 8. As it turns out, the fits by Eqs. (7) and (8) provide an equally good description for data of the long-time self-diffusion coefficient that span five orders of magnitude. The critical exponent $\gamma$ is very near to the predictions of MCT. Similar results have been found in studies where the application of MCT was tested for hard sphere systems by experiment [29] and simulation [27]. Finally, in other approaches, the dependence of $D_S^L$ on concentration has been described by a stretched exponential form [30] but they will not be discussed further here.

### 3. Collective diffusion coefficient

As made clear from the above discussion, the single particle diffusion motion is appreciably hindered by both direct and indirect interactions and hence the long-time self-diffusion coefficient $D_S^L(\phi)$ exhibits a strong dependence (decrease) on volume fraction. On the contrary, in hard sphere suspensions where sort-ranged repulsive interactions are important, the collective diffusion coefficient $D_C$, which describes the decay of long-wavelength density fluctuations, is found to depend only weakly on volume fraction [1]. The collective diffusion coefficient at short times has been theoretically determined by Batchelor [22(a)] taking into account of all hydrodynamic interactions in the limit of low volume fractions and is given by,

$$\frac{D_C^S(\phi)}{D_0} = 1.454\phi - 0.45\phi^2 + O(\phi^3). \qquad (9)$$

Experimental test of the above relation using a hard sphere colloidal suspension showed an underestimation of the linear term [31]. The dashed lines in Fig. 6 represent best (linear in $\phi$) fits to the experimental data for the collective diffusion coefficient using Eq. (9). The slope coefficients for the data at 20 $^\circ$C and 37 $^\circ$C were found 1.55 and 1.95, respectively. The deviations from theoretical predictions may arise from the fact that crystallins are not ideal hard sphere systems and show high polydispersity as well as due to the extrapolation from relatively high volume fractions, in view of the lack of collective diffusion data at $\phi \rightarrow 0$. There are, however, two worth-noting points. First, the extrapolation of $D_C(\phi)$ at $\phi \rightarrow 0$ results in a value for $D_0$ similar to that obtained by the limiting value of the self-diffusion coefficient, as is also evident from the dashed lines at both temperatures in Fig. 6. These data are tabulated in Table I. Second, the slope of $D_C(\phi)$ at 20 $^\circ$C is lower than the corresponding slope of the data at 37 $^\circ$C. Considering that repulsive interactions lead to an increase of the slope while the opposite occurs for attractive interactions we conclude that net repulsive interactions are weaker at 20 °C in comparison to 37 °C. This is a reasonable result knowing that the Coulobic repulsive interactions are far less affected by temperature compared with attractive van der Waals interactions. The latter increase with decreasing temperature, resulting in a weakening of the net repulsive forces.

**Table I:** *Limiting values for $D_0$ as estimated from the collective (c) and the self-diffusion coefficient data. $R_{h,0}$ stands the corresponding hydrodynamic radii obtained from $D_0$ (s).*

| T (°C) | $D_0$ (c) [$10^{-11}$ m$^2$ s$^{-1}$] | $D_0$ (s) [$10^{-11}$ m$^2$ s$^{-1}$] | $R_{h,0}$ [nm] |
|---|---|---|---|
| 20 | 1.72 | 1.70 | 12.6 |
| 37 | 2.60 | 2.54 | 12.9 |

In general, the concentration dependence of the collective diffusion coefficient is written as $D(c) = D_0(1 + k_d c)$ where the parameter $k_d$ is related to the thermodynamic second virial coefficient, $A_2$, through the equation:

$$k_d = 2A_2 M_w - k_f - 2\upsilon_{sp} \qquad (10)$$

where $k_f$ is a frictional coefficient describing effects arising from hydrodynamic interactions and $\upsilon_{sp}$ is the specific molar volume of the solute. This equation shows explicitly that interparticle interactions in suspension can be determined by the concentration dependence of $D(c)$. These interactions involve solute-solvent, solute-solute, and solvent-solvent interactions. Solute-solvent interactions are determined by the sign and the magnitude of $A_2$ indicate how good is the solvent for the solute particles; $A_2 > 0$ suggest a good solvent, while $A_2 < 0$ suggests a poor one. Solute-solute interactions are determined by $k_d$ (Eq. 10); $k_d > 0$ implies repulsive interactions, while $k_d < 0$ implies attractive interactions among solute particles.

The second virial coefficient of α-crystallins suspensions has been determined at various temperatures in Ref. [32]. The ratio of $A_2$ at the two temperatures of interest was found $A_2^{37}/A_2^{20} \approx 1.5$. This value is relatively close to the ratio of the $k_d$ values of the present study $k_d^{37}/k_d^{20} \approx 1.3$. The small difference can originate from the fact that in this work we study homogenate mixtures containing all species of crystallins. The Flory temperature, i.e. the point at which $A_2$ will turn negative was estimated at about –2 °C for α-crystallins suspensions [32]. The fact that $k_d$ as determined by the $D_C$ vs. $c$ data tends to a negative value faster than $A_2$ [32] is indicative of the specific role of the γ-crystallin species in the homogenate suspensions whose existence and attractive interactions lead to the well known cold cataract effect (or liquid-liquid phase separation of the lens cytoplasm) at a particular temperature lower than the physiological one.

## V. SUMMARY AND CONCLUDING REMARKS

A dynamic light scattering investigation on bovine lens homogenate protein dispersions has been carried out. The dependence of the diffusion modes and interparticle interactions on protein concentration and temperature has been the main focus of the present paper. This work has been driven by the need to understand the complex nature of the scattered light pattern emerging in studies of the intact mammalian lenses, which hampers a reasonable interpretation. Similar studies have been undertaken in the past in lens protein dispersion were isolated species of these proteins were studied, mainly the α-crystallin class. On the contrary, we have accomplished the first DLS study on lens homogenates including both the nucleus and the cortex of the lens thus maintaining the correct proportionality between the various classes of lens proteins. This is of particular importance since interactions between heterologous proteins have been considered decisive for the maintenance of lens transparency at high concentration [see for example 13(b) and references therein]. Specifically, while it is rather well established that repulsive interactions of Coulobic origin are responsible for the short-range order, which ultimately determines lens transparency at high concentration, there is growing evidence that attractive heterologous and homologous interactions might also contribute to this effect [13(b)].

Summarizing, in the present study two diffusion modes are observed even in the very dilute dispersions of lens homogenates. From detailed analysis and using information from existing biochemical studies these modes are considered to arise from the self-diffusion of isolated α-crystallin particles and high molecular weight assemblies of α-crystallins present in the lens nucleus. The self-diffusion modes show strong concentration dependence, exhibiting appreciable decrease with increasing volume fraction. These modes were identified with the long-time, self-diffusion modes and the volume fraction dependence of the relevant diffusion coefficients was discussed in the context of few existing theoretical approaches. No temperature dependence was observed for the long-time, self-diffusion modes.

With increasing concentration, changes in the intensity correlation functions are observed at about 75 mg ml$^{-1}$ signifying the visibility of the relaxation mode associated with the collective diffusion motions. Associating this with the log-time collective diffusion mode what we observe is in accordance with theoretical predictions that have shown that this mode cannot exist at low concentrations (i.e. for $\phi < 0.2$ for hard spheres) due to the weakening of the caging effect. The collective diffusion coefficient exhibits a mild volume fraction dependence, which becomes even weaker with decreasing temperature signifying a weakening of the net repulsive interactions. This temperature change is comparable with the corresponding change of the second virial coefficient measured in α-crystallins dispersions.

A comparison between experimental data of lens homogenate suspensions and intact bovine lenses has revealed significant changes between the time correlation functions at 37 $^{o}$C and 20 $^{o}$C. In the lens, these concern both changes of the amplitudes of fast and slow decaying modes as well as the appearance of new slow modes at 20 $^{o}$C, which can be associated with alterations in protein dynamic associate with the onset of cold cataract [7(b), 8(e)]. On the contrary, less drastic changes are observed in the protein dynamics of lens homogenate suspension for a high concentration (300 mg ml$^{-1}$) approaching that of the lens. The relative amplitude ratio between fast and slow modes remains practically unaffected while the changes in the time scale of dynamics are modest.

Closing this work, we would like to point out that the protein dispersions studied in this work represent a pragmatic biological system that lacks many idealities, which would render the interpretation of the results more straightforward and the comparison with theoretical approaches more direct. The complex nature of the suspensions under study is concerned with the fact that eye lens protein dispersions do not represent a true hard-sphere system, they show very high polydispersity, and contain three classes of proteins with disparate particles sizes and charges. While theories for hard-sphere, monodisperse systems abound, much less efforts have been directed in studies of more complex systems as described above. In any case, experimental studies in biological systems, such that studied here, can help illuminating new aspects in the physics of complex colloidal dispersion as well as to understand basic mechanisms of tissue functions and the routes to their degradation (lens cataract) under pathological conditions.

**Acknowledgments:** Financial support of the "ΠΕΝΕΔ-01/ΕΔ-559" project is gratefully acknowledged. "ΠΕΝΕΔ-01/ΕΔ-559" project is co-funded: 75% of public financing from the

*J. Chem. Phys, in press.*

European Union – European Social Fund and 25% of public financing from the Greek State – Ministry of Development – GSRT in the framework of the Operational Program "Competitiveness", Measure 8.3 – Community Support Framework 2000-2006. The authors thank Mrs. V. Petta for providing the data of the intact bovine lens shown in Fig. 1. SNY thanks Dr. G. Petekidis and Prof. M. Tokuyama for useful discussions on the subject of Section IV.